\documentclass{article}

\usepackage{arxiv}

\usepackage[utf8]{inputenc} 
\usepackage[T1]{fontenc}    
\usepackage{hyperref}       
\usepackage{url}            
\usepackage{booktabs}       
\usepackage{amsfonts}       
\usepackage{nicefrac}       
\usepackage{microtype}      
\usepackage{lipsum}
\usepackage{graphicx}
\usepackage{float}
\usepackage{amsmath}

\title{Deep Learning for Classifying and Characterizing Atmospheric Ducting within the Maritime Setting}

\author{
  Hilarie Sit \\
  School of Civil and Environmental Engineering\\
  Cornell University\\
  Ithaca, NY 14850 \\
  \texttt{hs764@cornell.edu} \\
   \And
 Christopher J. Earls\\
  School of Civil and Environmental Engineering\\
  Center for Applied Mathematics\\
  Cornell University\\
  Ithaca, NY 14850 \\
}

\begin{document}
\maketitle

\begin{abstract}
Real-time characterization of refractivity within the marine atmospheric boundary layer can provide valuable information that can potentially be used to mitigate the effects of atmospheric ducting on radar performance. Many duct characterization models are successful at predicting parameters from a specific refractivity profile associated with a given type of duct; however, the ability to classify, and then subsequently characterize, various duct types is an important step towards a more comprehensive prediction model. We introduce a two-step approach using deep learning to differentiate sparsely sampled propagation factor measurements collected under evaporation ducting conditions with those collected under surface-based duct conditions in order to subsequently estimate the appropriate refractivity parameters based on that differentiation. We show that this approach is not only accurate, but also efficient; thus providing a suitable method for real-time applications. 
\end{abstract}

 
\section{Introduction}
Propagation of electromagnetic (EM) signals through the marine atmospheric boundary layer (MABL) (\emph{i.e.} the lowest region of the troposphere above the ocean) can be heavily impacted by turbulent interactions at the air-sea interface. The MABL is characterized by turbulent fluxes of heat, moisture, and momentum between the ocean surface and the atmosphere that cause significant inhomogeneity in meteorological conditions within this region \cite{Sikora}. Spatial distributions of temperature, pressure, and humidity factor into the refractive index profile of the MABL; thus strongly influencing the behavior of EM waves propagating through this medium. For example, anomalous EM propagation can occur from a sharp decrease in the refractive index with respect to altitude above the ocean free surface \cite{Skolnik}. This specific type of anomalous propagation is called \emph{atmospheric ducting}, and depending on its effect on the refractive index profile, a duct can be categorized as an \emph{evaporation duct}, \emph{surface-based duct}, \emph{elevated duct}; combinations of these are also possible \cite{Karimian}. In ducting, the high refractive index gradient may create a \emph{trapping layer}, within which EM waves may be refracted back toward the ocean surface at a higher curvature than that of Earth's surface \cite{Skolnik}. Such behavior can negatively impact communication systems and navigational radars.

The specific case of radar performance motivates the current work: the goal of which is to arrive at a means for rapidly identifying and characterizing ducts within the MABL to allow radar operators on maritime vessels to gauge radar performance and to compensate for the effects of ducting in real-time. Such a capability requires a real-time estimation of the refractive conditions within the MABL. One approach for achieving this is to exploit the relationship between the refractive index and measurable meteorological conditions in order to directly calculate the refractive index \cite{Bean}. A variety of methods exist for collecting atmospheric data: thermodynamic sensors on both sides of a ship \cite{Babin}, radiosondes \cite{Dabberdt}, aircraft and buoy \cite{Brilouet}, \emph{etc}. Unfortunately, for the purposes of real-time duct characterization, direct measurements of meteorological conditions, such as in the foregoing, can be time-consuming, expensive, and data-sparse (\textit{i.e.} sampling is carried out at too low of a spatio-temporal resolution). More direct measurements using lidar \cite{Willitsford} and target-of-opportunity GPS signals \cite{Lowry} can also be used to estimate refractivity, but these methods have their own limitations for this particular problem (\textit{e.g.} the need for precise satellite alignment over the horizon and ideal weather conditions, respectively). Several alternative approaches to direct measurement involve the use of the radar transmitter, along with various measurements related to EM propagation.

\textit{Refractivity from clutter} (RFC) methods use radar clutter (\emph{i.e.} radio signals backscattered from the rough ocean free surface) to estimate the refractive index profile \cite{Karimian}. A common implementation of RFC involves a forward model to predict clutter returns under certain environmental conditions, followed by an inversion to recover parameters of the refractivity profile assumed to be governing the ducting in the MABL. A comprehensive history and review of RFC methods can be found in \cite{Karimian}. More recent methods for optimizing the RFC inversion include particle swarm optimization \cite{Wang}, proper orthogonal decomposition \cite{FountoulakisRFC}, opposition-based learning \cite{Yang}, and dynamic cuckoo search \cite{Zhang}. Additionally, deep learning has become a popular technique due to recent improvements in neural network architecture and training. \cite{Tepecik} use neural networks as part of a hybrid model to refine inversion predictions from genetic algorithms. \cite{Guo} show that deep neural networks can efficiently predict the refractive index parameters associated with evaporation ducts and surface-based ducts, outperforming certain specific genetic algorithms and radial basis function neural networks. \cite{Tang} similarly demonstrate the success of deep learning strategies for predicting surface-based duct refractive index parameters as compared with a certain instance of particle swarm optimization. Many RFC inversion approaches have focused on either evaporation duct or surface-based duct profiles; however inversion for combined evaporation and surface-based duct profiles has also been introduced in \cite{Douvenot} and \cite{WangCombine}.

\emph{Other EM-based techniques} that involve the direct transmission of EM energy into the MABL use sparsely measured EM propagation factors (PFs) to estimate the refractive index profile. In one method, collections of PFs are expressed as proper orthogonal modes and organized into a library that can be used for efficient real-time interpolation \cite{Fountoulakis}. Similarly, a low order basis of normal modes can be constructed from the sampled PFs to recast the multimodal inversion problem into a simplified optimization problem \cite{Gilles}. Other work utilizes the PFs directly, to obtain duct parameter predictions using artificial neural networks \cite{Sit} and Gaussian process regression \cite{Sit2}.

While inversion for duct parameters associated with a specific refractivity profile has achieved some success, there is a need for a real-time model that can differentiate between multiple types of ducting phenomena, so that subsequent duct characterization might be carried out. In the present paper, we introduce a two-step deep learning model to: 1) classify the duct; and 2) subsequently estimate the refractive index parameters corresponding to the identified class. We leverage the recent successes in the literature in applying deep learning models to predict evaporation duct and surface duct refractive index parameters and precede such analyses with a classification step.

The present paper is organized as follows: we begin, in Section 2, by describing our forward model and the creation of a pre-generated dataset that consists of sparsely sampled propagation factors; we then discuss model selection and deep neural network training in the context of the MABL duct classification and regression problems in Section 3; finally, we offer an analysis of the two-step model's performance on unseen, noise-contaminated data in Section 4.

\section{Forward Model}
Electromagnetic wave propagation within the MABL can be modelled using the time-independent Helmholtz wave equation \cite{Gilles}: 
\begin{equation} 
\frac{\partial^2 \varphi}{\partial x^2} + \frac{\partial^2 \varphi}{\partial z^2}  + k^2n(z)^2\varphi = 0
\end{equation}
where $\varphi$ is the horizontal polarization of the electric field, $x$ is the range, $z$ is the altitude, $k$ is the free-space wave number, and $n$ is the refractive index (\emph{i.e.} the ratio of EM propagation velocity in free space to that within the medium). To more efficiently estimate propagation for MABL duct characterization, the Helmholtz wave equation can be approximated using a parabolic equation \cite{Ozgun}; this is further discussed in section 2.3.

\subsection{Refractive Index}
Meteorological conditions within the MABL impact the propagation of EM waves by affecting the refractive index term in Eqn. 1. The refractive index profile can be related to atmospheric temperature, pressure and humidity in a relationship derived by \cite{Bean}. Within the standard atmosphere, the refractive index steadily decreases with altitude as a result of decreasing temperature, pressure, and humidity \cite{Ko}. Rays of EM energy are deflected as the refractive index decreases; thus, EM waves under standard propagation conditions bend slightly downward with a curvature double that of Earth's curvature \cite{Ko}. To better model the refractive conditions within the MABL, it is practical to define another quantity called the \textit{modified refractivity}, $M(z)=(n-1 +z/R_e) \cdot 10^6$, where $R_e$ is Earth's radius, to account for Earth's curvature and to amplify the small changes in the refractive index value \cite{Saeger}.

Sharp changes in temperature, pressure, and/or humidity gradients are not uncommon within the MABL, and can lead to anomalous propagation events such as \emph{ducting}, which occurs as a result of a sharp decrease in refractivity. The three primary duct classes are evaporation ducts, surface-base ducts, and elevated ducts. To characterize their effects on the refractivity within the MABL, we commonly use simplified modified refractivity models for each of the duct classes, in order to reduce the number of parameters we must search over to characterize the duct \cite{Saeger}. We additionally assume invariance of the modified refractivity profile with range along the path of transmission. Figure 1 shows the simplified modified refractivity profiles of three common ducts as well as the parameters of interest in this study, shown in red.

\begin{figure}[H]
\centering
\includegraphics[width=0.8\linewidth]{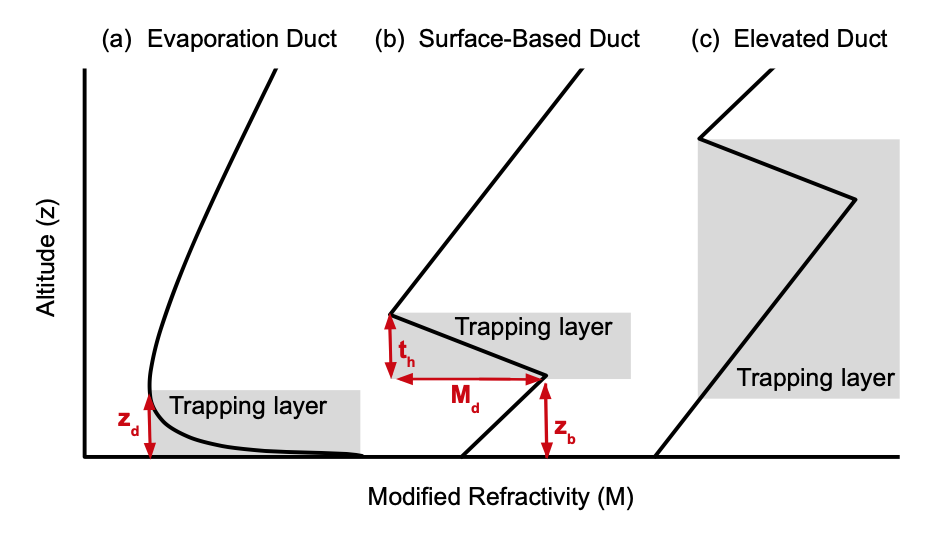}
\caption{Modified refractivity profiles of three common duct types: (a) evaporation, (b) surface-based, and (c) elevated}
\end{figure}

\subsubsection{Evaporation Duct}
The most prevalent ducts in the MABL are evaporation ducts \cite{Yardim}. Evaporation ducts form in the presence of a high humidity gradient caused by the sharp decrease in water vapor content, away from the ocean free surface \cite{Cook}. Models based on bulk measurements are commonly used to describe the refractivity profile of the MABL under evaporation ducting, because measuring refractivity close to the ocean surface can be difficult \cite{Cook}. The most popular bulk model is the \emph{Paulus-Jeske} (PJ) model and, for neutral conditions (\emph{i.e.} when the difference in temperature between the ocean surface and atmosphere is close to zero), the modified refractivity can be simplified as the following \cite{Karimian}:
\begin{equation}
M(z) = M_0 + c \left(z - z_d \ ln\left(\frac{z+z_0}{z_0} \right) \right) 
\end{equation}where $M_0$ is the surface refractivity, $c =  0.13\text{M-units/m}$ is the critical potential refractivity gradient, $z_0 = 0.00015 m$ is the aerodynamic surface roughness of the ocean, and $z_d$ is the duct height (Figure 1a). The value of surface refractivity has minimal effect on propagation and is fixed as $M_0 = 340\text{M-units}$ for all ducts within this study \cite{Vasudevan}. The duct height, defined as the altitude at which the gradient of the modified refractivity profile is zero, is an important parameter for evaporation ducts because it corresponds to their \emph{strength}: the greater the height, the stronger the duct and its ability to impact EM waves at lower frequencies \cite{Paulus}. The PJ model limits evaporation duct heights to be below 40m \cite{Babin}; thus, in this study, we consider $z_d$ between $2\text{m}$ and $40\text{m}$.

\subsubsection{Surface-Based Duct}
Although not as common as evaporation ducts, surface-based ducts can have a greater effect on radar efficacy: causing spurious increases in radar ranging, as well as clutter returns around the transmitter \cite{Karimian}. Surface-based ducts typically form within the MABL as a result of humidity and temperature inversions caused by warm and dry air moving over the cooler ocean surface \cite{Skolnik}. A tri-linear refractivity model, employing four parameters, can be used to instantiate a surface-based duct \cite{Gilles}:
\begin{equation}
    M(z) = \begin{cases} 
    M_0 + s_1z & z \leq z_b \\
    M(z_b) - \frac{M_d}{t_h}(z-z_b) & z_b \leq z \leq z_b + t_h \\
     M(z_b+t_h) - 0.118z & z_b + t_h \leq z
\end{cases}
\end{equation}
where $s_1$ is the initial slope of the refractivity, $z_b$ is the base height, $M_d$ is the M-deficit, and $t_h$ is the duct thickness (Figure 1b). Following \cite{Gilles}, we fix the initial slope at $s_1= 0.118 \text{M-units/m}$, and aligning closely to the ranges they considered, we consider $z_b$ between $1\text{m}$ and $30\text{m}$, $t_h$ between $1\text{m}$ and $30\text{m}$, and $M_d$ between $4\text{M-units}$ and $50\text{M-units}$. 

\subsubsection{Elevated Duct}
Elevated ducts form under a similar mechanism as surface-based ducts, and their effect on the modified refractivity profile can also be modeled as a tri-linear function (Figure 1c) \cite{Yardim}. Unlike surface-based ducts, elevated ducts do not come in contact with the ocean surface and are located at higher altitudes, typically between $600\text{m}$ to $3000\text{m}$ \cite{Dinc}. Due to their location in the atmosphere, elevated ducts do not influence radar systems in a way that is salient to the present study \cite{Karimian}; thus we do not treat them in the present work. We limit our study to evaporation ducts and surface-based ducts, although their combinations can be explored in future work.

\subsection{Problem Domain}
EM propagation from a transmitter antenna extends from the sea surface to infinity. Due to practical considerations and for computational feasibility, we consider a finite, rectangular 2D domain and apply the necessary boundary conditions to mimic the semi-infinite physical domain. In our domain, the range spans from $x = 0\text{km}$ (at the location of the transmitter antenna)  to $x = 50\text{km}$ with grid spacing of $40\text{m}$, and the altitude spans from $z = 0\text{m}$ (at the ocean surface) to $113\text{m}$ with grid spacing of $0.1\text{m}$.

As the altitude approaches infinity, EM energy from the transmitter antenna decays to zero. A \emph{Sommerfeld radiation condition} is typically used to achieve this behavior in the forward solver \cite{Ryan}. However, a truncated problem domain can cause artificial reflections in the solution, posing a challenge when implementing the upper boundary condition \cite{Nataf}. Our solution is to extend the upper boundary of the problem domain, and apply a Hanning window to smoothly decay the EM field; as suggested in the literature \cite{Ozgun}. At the lower boundary, the \emph{Leontovich surface impedance condition} is used to enforce continuity in the tangential components of the electric and magnetic fields; in order to do so, the ocean surface is modeled as a thin, flat, finite conductor \cite{Ryan}. In this study, the dielectric constant of the ocean surface aligns with thermodynamic conditions of the South China Sea during the SCSMEX experiments \cite{SCSMEX}: $100\%$ humidity at ocean surface, surface temperature of $29.7^{\circ}C$,  and ocean salinity of 35 ppt. The dielectric constant under these thermodynamic conditions can then be calculated using the semi-empirical \emph{Debye expression} \cite{Ryan}:
\begin{equation}
\epsilon(\omega) = \epsilon_{ir} + \frac{\epsilon_{0} - \epsilon_{ir}}{1-i\omega\tau} + \frac{i\sigma}{\omega\epsilon_{0}}
\end{equation}
where $\epsilon_{ir} = 4.9$ is the far-infrared dielectric constant of water, $\tau$ is the relaxation time, $\sigma$ is the ionic conductivity, and $\epsilon_0$ is the static dielectric constant of sea water.

\subsection{SSPE Solution}
In this study, surrogate data is generated using PETOOL \cite{Ozgun}, which utilizes the \emph{Fourier split-step parabolic equation} (SSPE) algorithm to solve the parabolic equation approximation of the Helmholtz equation, that has been modified by \cite{Gilles}. In SSPE, the initial field is specified at the location of the transmitter antenna (\emph{i.e.} left boundary in the problem domain) by using parameters of its antenna pattern. In this work, a Gaussian antenna pattern is specified using antenna height of $10\text{m}$, antenna angle of $0^{\circ}$, and radar signal frequency of 9.3 GHz. The initial signal is then propagated down range with fast Fourier transformations. To obtain the features of interest in this study, \emph{propagation factors} (PFs) (\emph{i.e.} electric field strength scaled with that observed in free space) are calculated using the SSPE solution \cite{Ozgun}: 
\begin{equation}
PF = 20log|u| + 10logx + 10log\lambda
\end{equation}where $u = \text{exp}(-ikx)\varphi(x, z)$ is the reduced amplitude function for the parabolic equation, and $\lambda$ is the free-space wavelength. In radar calculations, these propagation factors are strongly influenced by environmental effects and surface roughness \cite{Ryan}. It is pointed out that we purposely ignore surface roughness in the ocean free surface, as we are concerned with demonstrating the possibility of employing deep learning to classify, and subsequently characterize, EM ducts. More practical considerations could follow - in later research work.

\subsection{Sampling Path}
An unmanned aerial vehicle (UAV), outfitted with an RF receiver, could provide a practical way to obtain coverage measurements within the MABL. UAVs can be pre-programmed with a desired flight path above the ocean free surface. To conserve time to get these measurements for our prediction models, a linear flight path is considered in this study: inspired by \cite{Fountoulakis} who proposed a rocketsonde-receiver sampling system (RRSS) consisting of a X-band receiver attached to a solid rocket that is horizontally launched at a constant altitude. Using this latter approach, we can collect propagation factors down range from the radar within an order of minutes. In the present work, we acquire 250 equally spaced PF values sampled between 20km and 50km along the range direction from the transmitter, at an altitude of 6m above the ocean free surface. These sampled PFs, along with the duct classification and associated duct parameters, constitute an instance within our labeled surrogate dataset.

\section{Deep Learning}
In the present work, we show that a two-step deep learning model can accurately classify and characterize atmospheric ducting phenomena from sparsely sampled propagation factors. This two-step model consists of a classification model and associated regression models for evaporation ducts and surface-based ducts, as can be seen in Figure 2. PF arrays are passed, as inputs, to the classification model, and upon its subsequent prediction of duct type, the same PFs are then fed into the appropriate regression model to yield salient refractivity parameter estimates. We use deep neural networks for all our models and train them using supervised machine learning, to learn an approximate function that maps propagation factors (\textit{i.e.} inputs / feature vectors) to: 1) the duct type (\textit{i.e.} class) and 2) the associated collection of duct parameters (\textit{i.e.} regression labels).

\begin{figure}[H]
\centering
\includegraphics[width=\linewidth]{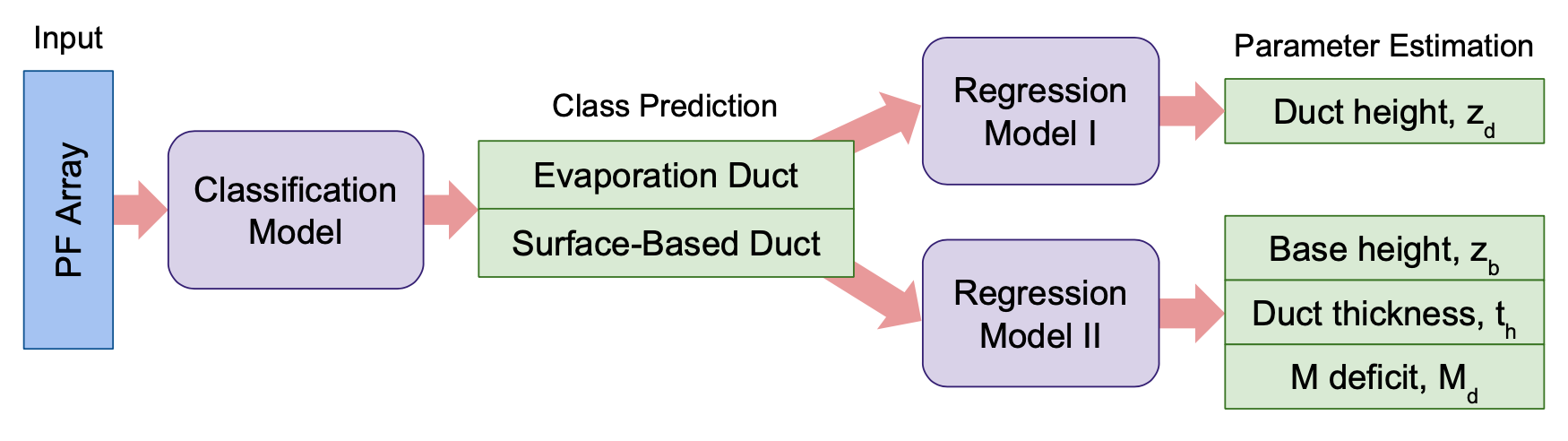}
\caption{Proposed two-step deep learning model}
\end{figure}

\subsection{Neural Network Architecture}
A \emph{deep learning} (DL) \emph{artificial neural network} (ANN) consists of multiple hidden layers of stacked \emph{perceptrons} that connect the input layer to the output layer, as shown in Figure 2. Within a single perceptron, commonly called a neuron within a neural network, inputs are multiplied by weighting factors, $w$, and summed with a scalar bias, $b$, such that a nonlinear activation function, $g$, can then be applied. The result is called the \emph{activation}: $a_i = g(x^T W + b_i)$. As universal function approximators, ANNs are able to approximate many nonlinear functions due to these nonlinear activation functions \cite{Shalev-Shwartz}. A common activation function for hidden layer neurons is the rectified linear unit (ReLU), $ ReLU(h) = max(0, h)$, which is computationally efficient and minimizes the vanishing gradient problem (by having a gradient of 0 for inactive units and 1 for active units) \cite{Nair}. In a neural network, the output of every neuron within the previous hidden layer becomes an input into the perceptrons within the subsequent layer.

The ``depth'' and ``width'' of the neural network are governed by the number of hidden layers and number of neurons per layer, respectively. These configurations are termed \emph{hyperparameters} of the model (\textit{i.e.} these define the training scheme and network architecture whose \emph{parameters}, the weights and biases, are \emph{learned} during training) \cite{Shalev-Shwartz}. Following the hidden layers, at the end of the network, is the output layer, where the number of neurons and activation function depends on the task. For example, in a classification task, the number of output neurons correspond to the number of classes to be discriminated between, and a softmax function is commonly used to return probabilities over each of the classes. On the other hand, in a regression task, the number of neurons correspond to the number of continuous, scalar parameters that are desired as labels; thus no activation function is applied to the output layer.

\begin{figure}[H]
\centering
\includegraphics[width=0.8\linewidth]{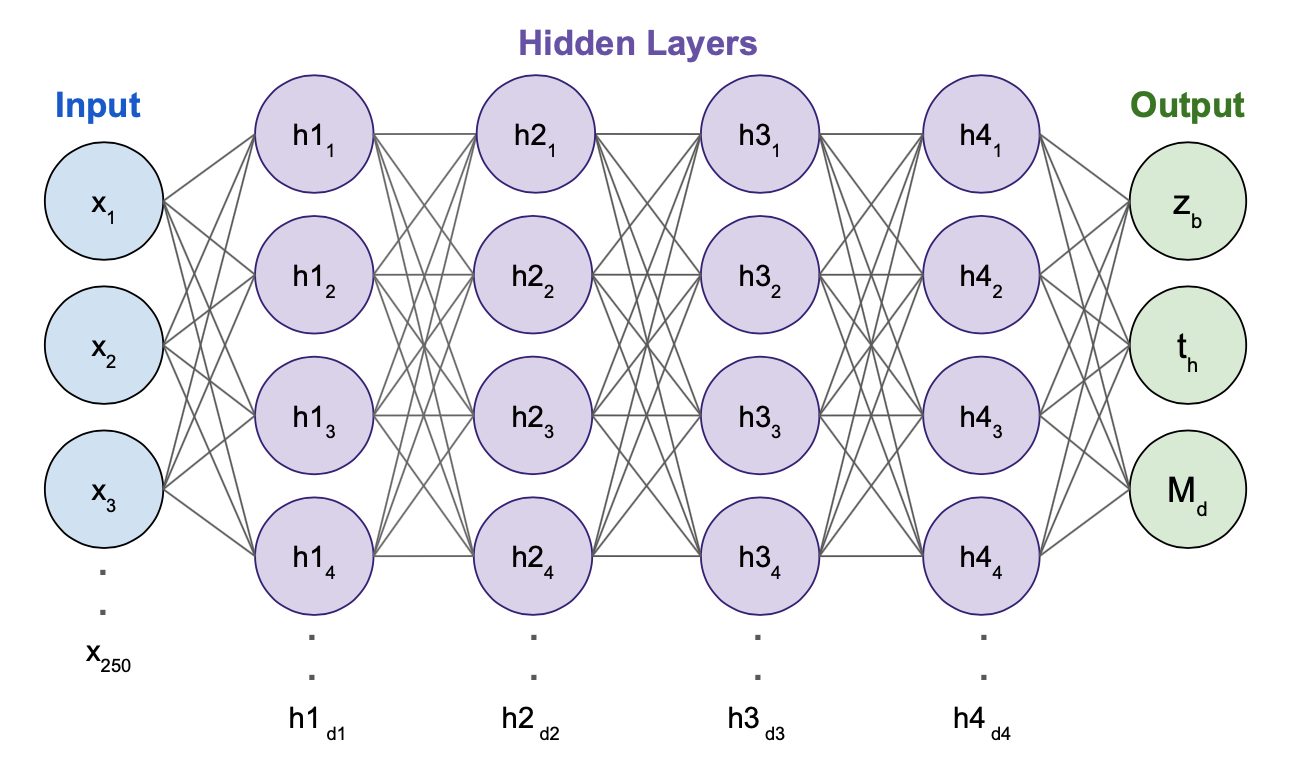}
\caption{Deep neural network for regression of surface-based duct parameters from propagation factors}
\end{figure}
 
\subsubsection{Regularization}
\textit{Overfitting} is a common problem for supervised machine learning algorithms, particularly within many physical contexts where data are limited. An overfitted model fails to generalize to unseen data, achieving low errors on the training set but furnishing large errors on the validation/testing set \cite{Shalev-Shwartz}. This phenomenon occurs if the model has too many parameters in comparison to the available training data, as is common in the case with deep neural networks. Several changes to the network architecture can help improve generalization of deep neural networks.

\emph{Batch normalization} (BN) is a popular technique used in deep neural networks to improve performance, accelerate training, and add regularization. In BN, inputs into the hidden layers of the network are standardized with their mean, $\mu$, and standard deviation, $\sigma$ \cite{Ioffe}:
\begin{equation} 
BN(u) = \gamma \cdot \frac{u-\mu}{\sigma} + \beta
\end{equation}where $\gamma$ and $\beta$ are learned parameters. Although the original paper applied BN before the activation function, in practice, it can also be applied after the activation function. The theory behind why BN is so effective is often disputed \cite{Bjorck}. It was originally suggested that BN reduces ``internal covariate shift'', which is the shift in the input distribution as parameters are tweaked \cite{Ioffe}. Through experimentation, \cite{Bjorck} deduced that BN prevents activations from blowing up, which allows for larger update steps, and in turn, speeds up convergence and improves regularization.
 
\emph{Dropout} is another powerful technique for improving model generalization at a low computational cost. In dropout, neurons and their connections are randomly removed from the network, with probability of $r$ (\emph{i.e} dropout rate, which is a hyperparameter) in each layer, by multiplying its activation by zero \cite{Srivastava}. By training the ANN with different activations zeroed out at every update, dropout forces many ``thinned'' sub-network models to learn the mapping function. The effect of dropout is similar to an ensemble of networks, which helps reduce variance by combining the predictions of multiple independent models \cite{Zhou}. When dropout is applied, model size is expanded by increasing network depth or width to help counter the decreased model capacity \cite{Goodfellow}. 

\subsection{Dataset}
Three mutually exclusive sets of data are commonly used to \emph{train}, \emph{validate}, and \emph{evaluate} supervised machine learning models to ensure that performance appropriately reflects the models' ability to make predictions on unseen data \cite{Shalev-Shwartz}. We generate three surrogate datasets using the forward model, as described in Section 2. In these datasets, single point consists of three pieces of information: feature (\emph{i.e.} array of 250 propagation factors), classification (\emph{i.e.} evaporation duct or surface-based duct), and label (\emph{i.e.} array of duct refractivity parameters). This could also be viewed as a class along with a vector of labels describing a given refractivity profile that is associated with points within a 250-dimensional \emph{feature space} comprising the sampled PFs. 

For the training set, \textit{Latin Hypercube sampling} (LHS) is employed to efficiently sample the multidimensional parameter space of our problem \cite{DouvenotSVM}. In LHS, the 1D sample space for every refractivity parameter is partitioned using a disjoint union of $N$ equally sized intervals. To instantiate a single point within the training dataset, for every refractivity parameter, an interval is randomly selected without replacement and a point is uniformly sampled from within that interval. This is repeated $N$ times to complete the training dataset. Using LHS, $N = 500$ evaporation duct environments and $N = 50,000$ surface-based duct environments are collected within the duct parameter ranges specified within Sections 2.1.1. and 2.1.2. 

For both the subsequent instantiation of validation and evaluation datasets, 50 evaporation duct environments and 5000 surface-based duct environments are uniformly sampled. To mimic experimental results with ``severe'' electronic sensor noise, the features in the test dataset are additionally contaminated with additive colored noise that has power spectral density of $1/|f|^\beta$, where $\beta$ corresponds to the inverse frequency power. In this study, both Gaussian white noise, $\beta = 0$, and pink noise, $\beta = 1$, are considered. The generated white noise is scaled by $\sigma = 0.1||x||_\infty$ (\textit{i.e.} 0.1 times the absolute value of the largest propagation factor); that scaling factor is divided by 1.6 for the generated pink noise in order to maintain consistent variance between white noise and pink noise.

An imbalanced dataset can have negative consequences in classification, causing the trained model to favor the more represented class. A common solution is to strengthen the role of the minority classes by either altering the loss function or balancing the class distribution (\emph{i.e.} up-sampling the smaller classes and down-sampling the larger classes) \cite{Menardi}. In this study, we randomly sample 50,000 points, with replacement, from the 500 point evaporation dataset to upsample the smaller class within the training set. We use the same method to correct the imbalance in the validation dataset. 

\subsection{Training}
During training, parameters of the ANN (\textit{i.e} $w$, $b$, $\gamma$, $\beta$) are randomly initialized and updated with an optimization algorithm, to improve a chosen loss function on the training dataset. Loss functions are selected based on the network application. For example, in binary classification, which involves a discrimination between two classes, a typical loss function employs the \emph{binary cross-entropy} \cite{Chollet}: 
\begin{equation}
L(y_i, \hat{y}_i) = - \frac{1}{N}[y_i \ log(\hat{y}_i) + (1 - y_i) \ log(1-\hat{y}_i)]
\end{equation}where $y_i$ is the label, $\hat{y}_i$ is the prediction, and N is the number of examples. Meanwhile, \emph{mean squared error} (MSE) is popular for regression tasks \cite{Chollet}:
\begin{equation}
L(y_i, \hat{y}_i) = - \frac{1}{N}(y_i -\hat{y}_i) ^2
\end{equation}
Variants of first-order gradient methods are commonly used to update parameters within a neural network. In first-order gradient methods, the gradient of the loss function is computed with respect to the network parameters in an approximate manner, such that parameters are adjusted along the directions of steepest descent in an attempt to reduce error. The magnitude of the parameter adjustments is dictated by the \emph{learning rate}. A trade-off exists in the selection of learning rate: a small learning rate can result in time-consuming and inefficient training phases, while a large learning rate can lead to divergence \cite{Shalev-Shwartz}. As a result of this trade off, it is common to use an adaptive learning rate that decreases throughout training. \emph{Adam optimization} is a stochastic first order method that utilizes estimation of the first and second moment of the gradient to calculate an adaptive learning rate \cite{Kingma}. Mini-batch based methods, which involve updating parameters utilizing only a subset of the data, rather than the complete dataset, can decrease update time and help improve generalization \cite{Goodfellow}. This subset from within the training dataset is called a \textit{mini-batch}; the size of which becomes yet another another hyperparameter of the network.

Adam optimization requires the calculation of partial derivatives of the loss function, taken with respect to the model parameters. For neural networks, an efficient way to calculate these gradients within the network is to use \textit{backpropagation}, or auto-differentiation \cite{LeCun}. Backpropagation employs information stored during the forward pass of the training data, in which the input is fed into the network to calculate activations, such that a backward pass can be made, in which loss function computed errors are fed back through the network to make model parameter adjustments in proportion to partial derivatives of the loss function (using stored information from the forward pass). The parameters are updated until a specified \emph{epoch}, which is defined as the stage at which the model has ``seen'' all the examples in the training dataset. We use Keras for this implementation \cite{Chollet}.

\subsection{Model Selection}
Model selection via direct, random search was performed to find optimal neural network hyperparameters for each of the three tasks (\emph{i.e.} classification as either surface or evaporation ducting, followed by regression to determine refractivity parameters). During a random search, a reasonable search space, based on problem specific considerations, is specified for each of the tunable hyperparmeters (\emph{e.g.} number of hidden layers, number of neurons per layer, dropout rate, batch size, and learning rate), and values are randomly selected within the space of admissible values, to generate multiple random neural network architectures \cite{Bergstra}. These network architectures are then trained and periodically evaluated using the validation dataset, and the high performing architectures are determined by the validation metric: the highest \emph{accuracy}, defined as the number of correct predictions over all predictions, for the classification task, and the lowest \emph{root mean squared error} (RMSE), defined as the square root of the average squared difference between the predicted value and true value, for the regression tasks.

Other hyperparameter optimization methods include \emph{manual search}, \emph{grid search}, and \emph{Bayesian optimization}. While Bayesian optimization can find optimal hyperparameters utilizing fewer training examples, both grid search and random search are embarrassingly parallelizable (\textit{i.e.} model architectures can be trained and evaluated independently); thus leading to computational advantages due to scaling over cores. \cite{Bergstra} showed that random search can achieve similar results to grid search while using less computation time. Thus, considering available computational resources (12‐cores of Intel Xeon E5 microprocessor with a clock speed of 2.7 GHz), parallelized random search was employed in the current study. 

In this study, the hyperparameter search space is defined using knowledge from previous literature, as well as prior experience. For each of the models, we consider between four and six hidden layers, with decreasing ranges for the number of hidden neurons in subsequent layers. We consider learning rates between 1e-3 and 1e-4, as well as appropriate dropout and batch size ranges for each task. For all our models, we use batch normalization and dropout to improve their ability to generalize to unseen data. Tables 1, 2, and 3 shows the five architecture and learning scheme combinations that achieved the best validation metric for the classification task, evaporation duct regression task, and surface-based duct regression task, respectively.

\begin{table}[H]
\caption {Hyperparameters corresponding to the top five classification models with the greatest validation accuracy}
\begin{center}
\begin{tabular}{ |c|p{1.15cm}|p{1.15cm}|p{1.15cm}|p{1.15cm}|p{1.15cm}|} 
\hline
Hyperparameters & Model 1 & Model 2 & Model 3 & Model 4 &  Model 5\\
\hline
hidden neurons & 922, 768, 418, 366, 237 & 935, 666, 597, 232, 305, 373 & 979, 684, 400, 206, 283 & 868, 641, 500, 351 & 832, 717, 571, 379, 247, 354 \\ 
learning rate (e-4) & 2.984& 4.831 & 8.245 & 7.279 & 5.442 \\
dropout & 0.2703 & 0.2137 & 0.3563 & 0.2878 & 0.2007 \\ 
batch size & 134 & 112 & 73 & 55 & 74 \\ 
\hline
training accuracy & 0.9930 & 0.9996 & 0.9958 & 0.9948 & 0.9997 \\ 
\textbf{validation accuracy} & \textbf{1} & \textbf{0.9999} & \textbf{0.9999} & \textbf{0.9999} & \textbf{0.9998} \\
\hline
\end{tabular}
\end{center}
\end{table}

\begin{table}[H]
\begin{center}
\caption {Hyperparameters corresponding to the top five evaporation duct regression models with the least validation loss}
\begin{tabular}{ |c|p{1.15cm}|p{1.15cm}|p{1.15cm}|p{1.15cm}|p{1.15cm}| } 
\hline
Hyperparameters & Model 1 & Model 2 & Model 3 & Model 4 &  Model 5\\
\hline
hidden neurons & 829, 603, 474, 284, 265, 319 & 833, 754, 499, 250, 288, 220 & 935, 719, 426, 248, 399 & 811, 616, 460, 323, 303 & 983, 713, 539, 395, 322 \\ 
learning rate  (e-4) & 5.442 & 9.708 & 7.279 & 9.894 & 6.822\\ 
dropout & 0.2007 & 0.3432 & 0.2878 & 0.2606 & 0.3325\\ 
batch size & 50 & 58 & 84 & 93 & 85\\ 
\hline
training loss & 0.07220 & 0.08977 & 0.06810 & 0.09150 & 0.08445 \\ 
validation loss & 0.04241 & 0.05730 & 0.05919 & 0.05978 & 0.05997 \\ 
\textbf{validation RMSE} & \textbf{0.2059} & \textbf{0.2394} & \textbf{0.2433} & \textbf{0.2445} & \textbf{0.2449} \\ 
\hline
\end{tabular}
\end{center}
\end{table}

\begin{table}[H]
\begin{center}
\caption {Hyperparameters corresponding to the top five surface-based duct regression models with the least validation loss}
\begin{tabular}{ |c|p{1.15cm}|p{1.15cm}|p{1.15cm}|p{1.15cm}|p{1.15cm}| } 
\hline
Hyperparameters & Model 1 & Model 2 & Model 3 & Model 4 &  Model 5\\
\hline
hidden neurons & 910, 745, 401, 377, 284, 344 & 863, 688, 507, 370, 370, 297 & 913, 740, 550, 392, 254 & 859, 638, 418, 364, 333, 249 & 901, 672, 494, 350, 218, 217\\ 
learning rate  (e-4) & 8.834 & 4.522 & 6.910 & 6.108 & 3.613\\ 
dropout & 0.265 & 0.3081 & 0.3226 & 0.2382 & 0.3151\\ 
batch size & 144 & 189 & 160 & 167 & 194 \\ 
\hline
training loss & 0.8504 & 0.9507 & 1.022 & 0.8433 & 1.066 \\ 
validation loss & 3.081 & 3.098 & 3.108 & 3.132 & 3.135 \\ 
\textbf{validation RMSE} & \textbf{1.755} & \textbf{1.760} & \textbf{1.763} & \textbf{1.770} & \textbf{1.771} \\ 
\hline
\end{tabular}
\end{center}
\end{table}

\subsection{Evaluation}
Architectures obtained from model selection in Tables 1, 2, and 3 are utilized in the \emph{two-step model}. We evaluate this two-step model on a separate test dataset, to gauge its performance on previously unseen data. We additionally assess the performance of the model on severely contaminated data (\textit{i.e.} test dataset contaminated with severe white and pink noise, as discussed in Section 3.2). 


Table 4 shows the accuracy of the classification models evaluated on the test dataset under different noise conditions. We also consider predictions from an \emph{ensemble}, which is a model that incorporates predictions from multiple models to reduce variance of the neural networks, potentially improving performance over a single model \cite{Zhou}. Many methods for ensembling models exist, such as averaging or weighted averaging the predictions, training another model to combine the predictions, averaging the weights of neural networks, \emph{etc}. In this study, we simply average the class probabilities from the top five models and calculate the argmax to obtain predictions. Subtleties in performance between the ensemble model and top model are highlighted by presenting the difference in the number of correct predictions. The predictions from the ensemble are then used to sift the data and send the examples into the correct regression model.

\begin{table}[H]
\begin{center}
\caption {Accuracy of classification models on test set during evaluation}
\begin{tabular}{|c|c|l|l|l|l|l|l|l|} 
\hline
Noise & Subset & Model 1 & Model 2 & Model 3 & Model 4 & Model 5 & Ensemble & Diff.* \\
\hline
none 
 & Evaporation & 98\% & 100\% & 100\% & 98\% & 100\% & 100\% & +1\\
 & Surface-based & 99.98\% & 99.90\% & 99.94\% & 99.88\% & 99.92\% &  99.96\% & -1\\
 \hline
white
 & Evaporation & 96\% & 100\% & 98\% & 98\% & 100\% & 100\% & +2 \\
 & Surface-based & 99.98\% & 99.88\% & 99.92\% & 99.88\% & 99.92\% & 99.94\% & -2\\
  \hline
pink 
 & Evaporation & 88\% & 96\% & 96\% & 96\% & 92\% & 96\% & +4\\
 & Surface-based & 99.86\% & 99.64\% & 99.84\% & 99.84\% & 99.82\% & 99.84\% & -1\\
  \hline
\end{tabular}
\\ *Difference in number of correct predictions of ensemble from best model
\end{center}
\end{table}

Tables 5 and 6 shows the \emph{RMSE} for the regression models as well as the ensemble model (averaging predictions from the top five models) corresponding to evaporation ducts and surface-based ducts, respectively. The percentage improvement of the ensembled predictions compared to the best model (\textit{i.e.} the negative percentage change: $-100\times(\text{RMSE}_{ensemble}-\text{RMSE}_{best})/\text{RMSE}_{best}$) is also presented for reference.

\begin{table}[H]
\begin{center}
\caption {RMSE of evaporation duct regression model on test set during evaluation}
\begin{tabular}{|c|c|l|l|l|l|l|l|l|l|} 
\hline
Parameter & Noise & Model 1 & Model 2 & Model 3 & Model 4  & Model 5  & Ensemble & \% Impr.* \\
\hline
$z_d$ (m) & none & 0.2721 & 0.2708 & 0.2496 & 0.2849 & 0.2967 & 0.1761 & 35.28\%\\
 & white & 0.2813 & 0.2822 & 0.2831 & 0.3610 & 0.3495 & 0.1999 & 28.94\%\\
 & pink & 0.4517 & 0.5024 & 0.4135 & 0.9953 & 0.4746 & 0.4267 & 5.545\%\\
\hline
\end{tabular}
\\ *Percent improvement of ensemble to best model
\end{center}
\end{table}

\begin{table}[H]
\begin{center}
\caption {RMSE of surface-based duct regression model on test set during evaluation}
\begin{tabular}{|c|c|l|l|l|l|l|l|l|l|} 
\hline
Parameter & Noise & Model 1 & Model 2 & Model 3 & Model 4  & Model 5  & Ensemble & \% Impr.* \\
\hline
$z_b$ (m) & none & 0.481 & 0.4391 & 0.4405 & 0.4167 & 0.4667 & 0.3841 & 20.14\%\\
 & white & 0.5282 & 0.5318 & 0.5623 & 0.513 & 0.5639 & 0.4759 & 9.904\% \\
 & pink & 2.395 & 2.568 & 2.622 & 2.465 & 2.672 & 2.364 & 1.308\% \\
\hline
$t_h$ (m) & none & 1.310 & 1.310 & 1.368 & 1.327 & 1.348 & 1.246 & 4.928\% \\
 & white & 2.105 & 2.105 & 2.083 & 2.020 & 2.125 & 1.940 & 7.825\% \\
 & pink & 3.602 & 3.522 & 3.545 & 3.503 & 3.606 & 3.278 & 8.997\%\\
\hline
$M_d$ (M-units) & none & 2.648 & 2.669 & 2.594 & 2.701 & 2.684 & 2.548 & 3.761\%\\
 & white & 3.74 & 3.659 & 3.729 & 3.666 & 3.739 & 3.502 & 6.364\% \\
 & pink & 4.182 & 4.15 & 4.157 & 4.102 & 4.096 & 3.911 & 6.484\% \\
  \hline
\end{tabular}
\\ *Percent improvement of ensemble to best model
\end{center}
\end{table}

The evaluation time for a single point was calculated for all three models by using python's \texttt{time\_process} to obtain the time elapsed in seconds. The times shown in Table 7 are averaged across all noise conditions and 5050 test examples. Times for both a single model (\textit{i.e.} the best model) and the ensemble are reported.

\begin{table}[H]
\begin{center}
\caption {Average Evaluation Time (s)}
\begin{tabular}{ |c|c|c|c|c| } 
\hline
& Classification & Regression & Total \\ 
\hline
Single & 5.550e-4 & 6.753e-4 & 1.231e-3 \\
Ensemble & 3.572e-3 & 4.600e-3 & 8.173e-3 \\
\hline
\end{tabular}
\end{center}
\end{table}

\subsection{Discussion}
Several observations can be made from the model selection results in Tables 1-3: validation losses are consistent across the top models within a task; models with four to six layers are represented, showing that a variety of model architectures can achieve comparable results;  finally, models in Tables 1 and 2 achieved lower validation loss than training loss, suggesting good regularization.

From results in Table 4, it is seen that deep neural networks are extremely efficient and accurate at classifying arrays of sparsely sampled propagation factors, as either measurements taken in the presence of evaporation ducting or surface-based ducting. Although trained solely on non-contaminated data, the classification models are robust to unseen data, even in the presence of severe white and pink noise. Under every noise condition, all five models achieved high accuracy, between $88\%$ - $100\%$ on evaporation duct PF measurements and $99.64\%$ - $99.98\%$ on surface-based duct PF measurements. This high performance can be difficult to improve; in fact, ensembling the models by averaging class probabilities generally yielded similar results to individual models. A notable $+4$ difference in the number of correct predictions can be seen for pink-noise contaminated evaporation duct measurements. For this set of data, the best model's performance of $88\%$ is lower than that of other models, suggesting its weakness in generalization. Ensembling neural networks can decrease outlier effects by reducing variances in network performance, and in this case, ensembling achieved a performance of $96\%$. 

For regression networks, ensembling the models improved RMSE to varying degrees, from $1.308\%$ to $35.28\%$. As can be seen in Tables 5 and 6, the percentage decrease in RMSE from ensembling can be high in certain cases and insignificant in others. Thus, for real-time applications, it is important to consider its benefits with the trade-off in computation time. From Table 7,  ensembling the networks increased the total computation time by a factor of approximately $6.64$; despite this, the two-step ensembled model can produce a prediction in under nine milliseconds.

\section{Conclusion}
The two-step deep learning model, used in conjunction with the sparse data collection process (\emph{i.e.} collecting 250 propagation factor measurements using a radiosonde programmed on linear flight path), can be effectively and efficiently deployed to distinguish duct types and characterize their effect on refractivity within the MABL in real-time. In the deep learning model, classification of ducts from sparsely sampled propagation factors achieved high accuracy on both the noise-free and color-noise contaminated test set. Subsequently, the regression models for both duct types achieved reasonable success relative to other deep learning models in the literature. Furthermore, we show that ensembling the top five regression models from hyperparameter search slightly improves predictions, although with the caveat of increased computation time. It is noted that the surrogate data used to train these models assumes certain environmental idealizations: horizontally invariant refractivity profiles, smooth sea surface, simplified modified refractivity profiles, \emph{etc}. Using more realistic training data, it is expected that the models would generalize well to more realistic cases. The observed accuracy and computational speed of the two-step deep learning model makes this a suitable option for real-time inference of ducting characteristics within the MABL. 

\section{Acknowledgments}
Datasets and code for this research are available at \url{https://github.com/nonlinearfun/deep-learning-em-ducting}. The authors gratefully acknowledge ONR Division 331 \& Dr. Steve Russell for the financial support of this work through grant N00014-19-1-2095.

\bibliographystyle{ieeetr}
\bibliography{paperbib}

\end{document}